\documentstyle[12pt]{article}
\begin{document}

\begin{center}
 { \bf Semiclassical Approximation for Periodic Potentials  } \\
\vspace{7mm}
\bf                     {   U. P. Sukhatme   }               \\
\vspace{2mm}
\it{  Department of Physics, University of Illinois at Chicago,
            Chicago, Illinois 60607, USA }  \\
\vspace{3mm}
\bf                     {   M. N. Sergeenko   }               \\
\vspace{2mm}
\it{ The National Academy of Sciences of Belarus, Institute of Physics, \\
                  Minsk 220072, Belarus ~ and \\ }
\it{  Department of Physics, University of Illinois at Chicago,
                      Chicago, Illinois 60607, USA }
\end{center}
\vspace{2mm}

\begin{abstract}
We derive the semiclassical WKB quantization condition for obtaining
the energy band edges of periodic potentials. The derivation is based
on an approach which is much simpler than the usual method of
interpolating with linear potentials in the regions of the classical
turning points. The band structure of several periodic potentials is
computed using our semiclassical quantization condition. \\ 

\noindent PACS number(s): 31.15.Gy, 31.70.Ks \end{abstract}

\section { 1. Introduction }

The study of periodic potentials is of both physical and mathematical
interest. For instance, in condensed matter physics, knowledge of
the existence and locations of band edges and band gaps of periodic
potentials is very important for determining many physical
properties. Although many general mathematical properties of the
eigenstates are known \cite{mw}, unfortunately, even in one
dimension, there are very few analytically solvable periodic
potential problems.

One of very successful methods in numerous applications in physics
and mathematics is the semiclassical approach. The semiclassical WKB
approximation for one dimensional potentials with two classical
turning points is discussed in most quantum mechanics textbooks
\cite{text}. It was originally proposed for obtaining approximate
eigenvalues in the limiting case of large quantum numbers. It has
been successfully used for many years to determine eigenvalues and to
compute barrier tunneling probabilities. The analytic properties of
the WKB approximation have been studied in detail from a purely
mathematical point of view, and the accuracy of the method has been
tested by comparison between analytic and numerical results
\cite{Dunh,Krei,SeSK}. There has been a special surge of interest
in recent years due to the development of the supersymmetric WKB
method \cite{Com,Sukh,Rep}.

In this paper, we generalize the standard WKB approach to treat
periodic potentials. We derive a new semiclassical quantization
condition whose solutions give the energy band edges of periodic
potentials. Our derivation is considerably simpler than the standard
approach which makes use of connection formulas \cite{Lang} to match
the WKB solutions in classically allowed regions with the solutions
in classically forbidden regions. In the usual approach, one makes a
linear approximation to the potential in the regions around the
classical turning points where the semiclassical first order WKB wave
function diverges. Although the connection formulas resulting from
the usual approach turn out to be simple enough, the derivation is
quite tedious. Our simpler approach effectively amounts to matching
the values and derivatives of the zeroth order WKB wave function
which is non-divergent at the classical turning points
\cite{SeAsym,Pag}.

In Sec. 2, we describe and justify our simpler approach by first
re-deriving the standard WKB quantization condition for potentials on
the infinite real line with two classical turning points. The
extension to periodic potentials of period $L$ is given in Sec. 3,
the main result being the quantization condition Eq. (\ref{qperi}).
To the best of our knowledge, this result and our method of
derivation have not been previously discussed, even though the
semiclassical approximation and periodic potentials have both been
studied for many years. In order to check the accuracy of the
semiclassical approach, we consider applications to the well-studied
class of Lam\'{e} potentials $V(x) = ma(a+1)\,{\rm sn}^2(x,m)$, where
${\rm sn}(x,m)$ is a Jacobi elliptic function \cite{gr}. This is a
good choice, since it is one of the very few potentials for which the
band edges are analytically known for integer values of $a$
\cite{ar,df,sk}. A comparison of band edge energies obtained from our
WKB quantization condition with exact results is given in Sec. 4. The
limitations and successes of the WKB approach for periodic potentials
are discussed.  \vspace{4mm}

\section { Simpler Derivation of the Usual WKB Condition }

In this section, we look at the standard situation of a potential on
the entire real line, which has two classical turning points $x_L$
and $x_R$ given by $V(x)=E$ for any choice of energy $E$. From now
on, for simplicity, we restrict our attention to symmetric potentials
$V(x)=V(-x)$. For this case, $x_L=-x_R$, and it is sufficient to just
look at the half line $x>0$, since the eigenfunctions will be
necessarily symmetric or antisymmetric. To derive the WKB
quantization condition we have to connect the solution in the
classically allowed region with the solution in the classically
forbidden region. For the symmetric case, the zeroth order WKB
approximation to the wave function for $x>0$ is

\begin{equation}
\psi_I^{(0)}(x) = A\cos[\chi(x)-\chi(0)]  \label{cla}
\end{equation}
in the classically allowed region $x_L\le x\le x_R$ and

\begin{equation}
\psi_{II}^{(0)}(x) = Be^{-\chi(x)+\chi(x_R)}   \label{ncla}
\end{equation}
in the classically forbidden region $x>x_R$. We are using the notation

\begin{equation}
\chi(x) \equiv \frac 1\hbar\int^x p(x,E)\,dx, \label{chiz}
\end{equation}
where $p(x,E) =\sqrt {2m|E-V(x)|}$ is the generalized momentum.
Matching the wave functions $\psi_I^{(0)}(x)$ and
$\psi_{II}^{(0)}(x)$ and their first derivatives at $x_R$ gives two
equations

\begin{equation}
 A\cos[\chi(x_R)-\chi(0)] = B, \label{e1}
\end{equation}

\begin{equation}
 A\sin[\chi(x_R)-\chi(0)] = B, \label{de1}
\end{equation}
which yield $\tan[\chi(x_R)-\chi(0)] = 1$, or

\begin{equation}
\frac 1\hbar\int_0^{x_R}p(x,E)dx=\frac 14\pi,\frac 54\pi,\frac
94\pi,\ldots  \label{q1} \end{equation}

Similarly for the antisymmetric case, the zeroth order WKB
approximation to the wave function for $x>0$ is $\psi_I^{(0)}(x)=
A\sin[\chi(x)-\chi(0)]$ in the classically allowed region $x_L\le
x\le x_R$ and $\psi_{II}^{(0)}(x) = Be^{-\chi(x)+\chi(x_R)}$ in the
classically forbidden region $x>x_R$. Matching these wave functions
and their first derivatives at $x_R$ now gives
$\tan[\chi(x_R)-\chi(0)] =-1$, or

\begin{equation}
\frac 1\hbar\int_0^{x_R}p(x,E)dx =
\frac 34\pi,\frac 74\pi,\frac{11}4\pi,\ldots \label{q2}
\end{equation}
Combining Eqs. (\ref{q1}) and (\ref{q2}), the quantization condition
is

\begin{equation}
\frac 1\hbar\int_0^{x_R}p(x,E) dx =\frac\pi 2\left(n+\frac 12\right),
~~n=0,1,2,\ldots    \label{usual}
\end{equation}
which is the usual WKB result for a symmetric potential \cite{text}.
The derivation given above is evidently much simpler than the usual
textbook approach for deriving connection formulas. However, some
comments, explanation and justification of the method used is needed
for several points. The usual approach makes use of first order WKB
wave functions $\psi^{(1)}(x)=\psi^{(0)}(x)/\sqrt{p(x)}$, which
diverge at the classical turning points. Although this divergence is
understandable in the classical limit, since a classical particle has
zero speed at the turning points, it is certainly not present in a
full quantum mechanical treatment. Since $\psi^{(1)}(x)$ is singular,
it is necessary to resort to connection formulas and somewhat tricky
matching of the wave function $\psi^{(1)}(x)$ and its first
derivative \cite{Lang}, that eventually yields the well known WKB
quantization condition Eq. (\ref{usual}).

Why our simple procedure for matching $\psi^{(0)}(x)$ is justified?
Clearly, the correct approach is neither to match $\psi^{(0)}(x)$ nor
$\psi^{(1)}(x)$, but to keep a sufficient number of higher order
contributions in $\hbar$, so that the resulting wave function is
non-divergent \cite{Pag}. This has to be the case, since there is no
divergence in the full wave function. A simple way in which the
divergence gets tamed is for the WKB wave function to have the form
$\psi^{WKB}(x)=\psi^{(0)}(x)/[\sqrt{p(x)}+\hbar f(x,\hbar)]$, where
$f(x,\hbar)$ is an analytic function of $x$ and $\hbar$. It is easy
to check that requiring $\psi^{WKB}(x)$ and its derivatives to be
continuous amounts to our procedure of matching the value and slope
of $\psi^{(0)}(x)$ at the classical turning point $x_R$, which
justifies our simple approach.  \vspace{4mm}

\section { Generalization to Periodic Potentials }

For a potential with period $L$, one is seeking solutions of the
Schr\"odinger's equation subject to the Bloch condition

\begin{equation} \label{1}
\psi (x) = e^{ikL}\ \psi(x+L)\, ,
\end{equation}
where $k$ denotes the crystal momentum. The spectrum shows energy
bands whose edges correspond to $kL=0,\pi$, that is the wave
functions at the band edges satisfy $\psi(x) = \pm\psi(x+L)$. For
periodic potentials, the band edge energies and wave functions are
often called eigenvalues and eigenfunctions, and we will also use
this terminology. A general property of the eigenfunctions for a
potential with period $L$ is the oscillation theorem \cite{mw} which
states that the band edge wave functions arranged in order of
increasing energy have periods $L, 2L, 2L, L, L, 2L, 2L, \ldots$.

For any periodic potential, it is sufficient to consider just one
period of width $L$, say the interval $[-\frac L2,\frac L2]$. In this
paper, we are discussing analytic potentials with $V_{min}$ and
$V_{max}$ as the minimum and maximum values. Further, we are only
looking at symmetric potentials $V(x) = V(-x)$, which necessarily
makes $x=0$ a maximum or minimum. Let us take the origin $x=0$ to be
at $V_{min}$. The eigenfunctions will either be symmetric (S) or
antisymmetric (A) about $x=0$. Furthermore, it is easy to see that
the potential is also symmetric about $x=\frac{L}{2}$, since
periodicity and symmetry about $x=0$ imply that $V(x+\frac{L}{2}) =
V(-x+\frac{L}{2})$. Consequently, $x=\frac{L}{2}$ is an extremum, and
the eigenfunctions are necessarily symmetric or antisymmetric about
$x=\frac{L}{2}$. Clearly, there are four types of eigenfunctions
(S,S), (A,S), (S,A), (A,A), where the first letter denotes symmetry
or antisymmetry about the origin, and the second letter denotes
symmetry or antisymmetry about the point $x=\frac{L}{2}$. Note that
(S,S) and (A,A) wave functions have period $L$, whereas (A,S) and
(S,A) wave functions have period $2L$. The ground state, being
nodeless, is of type (S,S).

To obtain the WKB quantization condition for a periodic potential,
consider any energy $E$ which gives two classical turning points
$x_L$ and $x_R$ in the interval $[-\frac L2,\frac L2]$. Since we are
looking at symmetric potentials, clearly $x_L=-x_R$, and one only
needs to look at the half-interval $[0,\frac L2]$. Note that the
region $|x|\le x_R$ is classically allowed (region I) whereas the
region $x_R<|x|\le\frac L2$ is classically forbidden (region II).
Let us now derive the WKB quantization condition corresponding to the
four eigenfunction types.

For the (S,S) case, the zeroth order WKB approximation to the wave
function is given by Eq. (\ref{cla}) in region I and

\begin{equation}
\psi_{II}^{(0)}(x)= B\cosh[-\chi(x) +\chi(L/2)] \label{nclap}
\end{equation}
in region II, where $\chi(x)$ is defined by Eq. (\ref{chiz}). Matching
the wave functions $\psi_I^{(0)}(x)$ and $\psi_{II}^{(0)}(x)$ and their
first derivatives at $x_R$ gives

\begin{equation}
A\cos[\chi(x_R)-\chi(0)]\! =\! B\cosh[-\chi(x_R)+\chi(L/2)],
\label{e2} \end{equation}

\begin{equation}
A\sin[\chi(x_R)-\chi(0)]\! = \!B\sinh[-\chi(x_R)+\chi(L/2)],
\label{d2} \end{equation}
which, on division, give the equation

\begin{equation}
\tan\,\left[\frac 1\hbar\int^{x_R}_0 p(x,E)\,dx \right]
= \tanh\,\left[\frac 1\hbar\int^{\frac L2}_{x_R} p(x,E)\,dx\right].
\label{qq1} \end{equation}

Similarly, for the (A,S), (S,A) and (A,A) cases, we take zeroth order
WKB wave functions with appropriate symmetry

\begin{eqnarray}
(A,S):\hspace{15mm}
\psi_I^{(0)}(x) = A\sin[\chi(x)-\chi(0)], \nonumber \\
\psi_{II}^{(0)}(x) = B\cosh[-\chi(x) +\chi(L/2)],  \nonumber
\end{eqnarray}

\begin{eqnarray}
(S,A):\hspace{15mm}
\psi_I^{(0)}(x) = A\cos[\chi(x)-\chi(0)], \nonumber \\
\psi_{II}^{(0)}(x) = B\sinh[-\chi(x) +\chi(L/2)], \nonumber
\end{eqnarray}

\begin{eqnarray}
(A,A):\hspace{15mm}
\psi_I^{(0)}(x) = A\sin[\chi(x)-\chi(0)], \nonumber \\
\psi_{II}^{(0)}(x) = B\sinh[-\chi(x) +\chi(L/2)], \nonumber
\end{eqnarray}
and match wave functions and their first derivatives at $x_R$.
Combining all results gives

\begin{equation}
\tan\,\left[\frac 1\hbar\int^{x_R}_0 p(x,E)\,dx \right] =
\pm \tanh\,\left[\frac 1\hbar\int^{\frac L2}_{x_R} p(x,E)\,dx\right],
\label{qp1}  \end{equation}

\begin{equation}
\tan\,\left[\frac 1\hbar\int^{x_R}_0 p(x,E)\,dx \right]
= \pm\coth\,\left[\frac 1\hbar\int^{\frac L2}_{x_R} p(x,E)\,dx\right].
\label{qp2} \end{equation}
Taking into account the periodicity of $\tan\chi(x)$, we can write a
single combined equation

\begin{equation}
\frac 1\hbar\int^{x_R}_0 p(x,E)\,dx = \frac\pi 2n \pm\arctan\left[
\tanh\,\left(\frac 1\hbar\int^{\frac L2}_{x_R} 
p(x,E)\,dx\right)\right].  \label{qperi}
\end{equation}

This is our final semiclassical WKB quantization condition for
symmetric periodic potentials of period $L$, the solutions to which
are the WKB band edge energies $E_n^{WKB}$. The quantum number $n$ takes
on non-negative integer values which keep the right hand side of Eq.
(\ref{qperi}) positive. Eq. (\ref{qperi}) is a generalization of the
usual two classical turning points quantization condition of Eq.
(\ref{usual}), which is found in quantum mechanics textbooks. This
can be readily established, since when $L \rightarrow \infty$,
$\tanh\,\left[\frac 1\hbar\int^{\frac L2}_{x_R}p(x,E)\,dx\right]$
tends to unity, and one gets Eq. (\ref{usual}).  \vspace{4mm}

\section { Applications and Discussion }

In order to study the accuracy of band edges resulting from the WKB
quantization condition derived in the previous section, we consider
several examples. These are selected from the class of Lam\'e
potentials

\begin{equation}
V(x) = ma(a+1)\,{\rm sn}^2(x,m), \label{sn}
\end{equation}
where the Jacobi elliptic function ${\rm sn}(x,m)$ has period
$4K(m)$. The potentials have a period $L = 2K(m).$ It is well known
that for any integer value $a = 1,2,3,\ldots,$ the corresponding
Lam\'e potential has $2a+1$ band edges corresponding to $a$ bound
bands followed by a continuum. All band edge eigenfunctions are
analytically known \cite{ar,df,sk}.

The Lam\'e potential given in Eq. (\ref{sn}) has $V_{min}=0$ and
$V_{max}=ma(a+1).$ A comparison of exact, analytically available band
edge energies with results obtained from the WKB quantization
condition applied to a variety of Lam\'e potentials is shown in Table
1.

The quantum umber $n$ and the symmetry of the wave functions about
the points $x=0$ and $x=\frac L2$ are also tabulated. The results are
consistent with the pattern required by the oscillation theorem for
periodic potentials \cite{mw}.  Note that for $m=1$, the period of
the Lam\'e potentials becomes infinite and all finite bands reduce to
zero width. This feature, which is evident for the exact energies of
the potential $V = 12\,{\rm sn}^2(x,1)$ considered in Table 1, is also
reproduced in the WKB approach. Also, in Table 1, it should be noted
that some exactly known band edges occur above $V_{max}.$ For
example, choosing $a=2, m=0.5$, one has the the potential $V(x)=
3\,{\rm sn}^2 (x,0.5).$ It has two energy bands ranging from 1.27 to
1.50 and from 3.0 to 4.5, with a continuum above 4.73. Clearly, the
band edges above $V_{max}=3$ cannot be obtained by the semiclassical
WKB method, since there are no classical turning points for $E >
V_{max}$. This is a limitation of the semiclassical approach.
However, for $V_{min}\le E\le V_{max},$ it is apparent from Table 1
that results obtained from our WKB quantization condition are in
modest agreement with exact band edge eigenenergies.  \vspace{3mm}

{\bf Acknowledgements.} It is a pleasure to acknowledge partial
financial support from the U.S. Department of Energy and the
Belarusian Fund for Fundamental Researches.

\newpage
\vspace{12mm}
{ Table 1: Exact and WKB band edge energies.}

\bigskip
\begin{tabular}{cccccc}
\hline
\hline
Potential & $~E_n^{\rm exact}~$ & $~E_n^{\rm WKB}~$ & $~n~$ &
Symmetry \\
\hline
\hline
$3\,{\rm sn}^2(x,0.5)$
& 1.27 & 1.34 & 0 & S,S \\
& 1.50 & 1.96 & 1 & A,S \\
& 3.00 & 2.81 & 1 & S,A \\
& 4.50 & - & - & - \\
& 4.73 & - & - & - \\
\hline
\hline
$6\,{\rm sn}^2(x,0.5)$
& 2.05 & 2.19 & 0 & S,S \\
& 2.13 & 2.35 & 1 & A,S \\
& 5.05 & 4.95 & 1 & S,A \\
& 6.00 & - & - & - \\
& 6.95 & - & - & - \\
& 9.87 & - & - & - \\
& 9.95 & - & - & - \\
\hline
\hline
$9.6\,{\rm sn}^2(x,0.8)$ & 2.68 & 2.87 & 0 & S,S \\
& 2.68 & 2.87 & 1 & A,S \\
& 7.04 & 7.10 & 1 & S,A \\
& 7.20 & 7.49 & 2 & A,A \\
& 9.32 & 9.12 & 2 & S,S \\
& 10.52 & - & - & - \\
& 10.96 & - & - & - \\
\hline
\hline
$12\,{\rm sn}^2(x,1)$ & 3.00 & 3.21 & 0 & S,S \\
& 3.00 & 3.21 & 1 & A,S \\
& 8.00 & 8.14 & 1 & S,A \\
& 8.00 & 8.14 & 2 & A,A \\
& 11.00 & 11.07 & 2 & S,S \\
& 11.00 & 11.07 & 3 & A,S \\
& 12.00 & 12.07 & 3 & S,A \\
\hline
\hline
\end{tabular}

\newpage
\oddsidemargin 0in

\end{document}